\documentclass[reprint,longbibliography,pre,amsmath,amssymb,amsfont, url]{revtex4-2}

\usepackage[colorlinks, allcolors=blue]{hyperref}
\usepackage{graphicx} 
\usepackage[inline]{enumitem}
\usepackage{eurosym}
\usepackage[cal=dutchcal]{mathalfa}
\usepackage{csquotes}
\usepackage{mathtools}

\usepackage{tikz}
\usetikzlibrary{arrows, fit}

\newcommand{\myquote}[1]{\textit{\textquote{#1}}}
\newcommand*\diff{\mathop{}\!\mathrm{d}}
\newcommand*\vx{\textbf x}
\newcommand*\W{\mathcal W}
\newcommand*\V{\mathcal V}

\newcommand{\law}[2]{
	\vspace{0.25cm}
	\noindent
	\textbf{#1}: 
	\textit{#2}
	\vspace{0.2cm}
}

\begin{document}
	\author{D. Lairez}
	\email{didier.lairez@polytechnique.edu}
	\affiliation{Laboratoire des solides irradi\'es, \'Ecole polytechnique,  CEA, CNRS, IPP,
		%Institut Polytechnique de Paris, 
		91128 Palaiseau, France}
	\title{Thermostatistics, information, subjectivity, why is this association so disturbing?}
	\date{\today}
	
	\begin{abstract}
		Although information theory resolves inconsistencies (known under the form of famous enigmas) of the traditional approach of thermostatistics, its place in the corresponding literature is not what it deserves.
		This article supports the idea that this is mainly due to epistemological rather than scientific reasons: the subjectivity introduced into physics is perceived as a problem. Here is an attempt to expose and clarify where exactly this subjectivity lies: in the representation of the reality and in probabilistic inference. 
		Two aspects which have been integrated into the practice of science for a long time and which should no longer frighten anyone, but which become explicit with information theory.
	\end{abstract}
	
	\maketitle

\section{Introduction}	

With great success, statistical mechanics provides \myquote{the rational foundations of thermodynamics} (Gibbs\,\cite{Gibbs_1902}) which thus becomes thermostatistics.
However,
\myquote{[it] is notorious for conceptual problems to which it is difficult to give a convincing answer} (Penrose\,\cite{Penrose_1979}).
Since the origin, these problems are illustrated by famous enigmas which do not prevent the theory from advancing, but are like pebbles in the shoe. 
I think about the two Gibbs' paradoxes (related to the mixing of two volumes of gas), the Poincaré-Zermelo paradox (related to the recurrence of dynamical systems), the Loschmidt's paradox (related to the reversibility of the equations of mechanics) and its demonic version, 
the Maxwell's demon (a ratchet-pawl mechanism at the scale of particles).
These enigmas all have one thing in common: they are all concerned with the second law of thermodynamics and entropy, the concept that was invented by Clausius\,\cite{Clausius_1879} to account for the irreversibility of energy exchanges, linked by  Boltzmann\,\,\cite{Boltzmann_Lectures}, Planck\,\cite{Planck_1914} and Gibbs\,\cite{Gibbs_1902} to probabilities and finally \textquote{enlightened} by Shannon's information theory\,\cite{Shannon_1948}. 

\textquote{Enlightened} is wishful thinking because, although information theory resolves the inconsistencies raised by these enigmas (as will be shown), this contribution is far from being unanimously recognized.
In recent textbooks, apart from one exception\,\cite{Ben-Naim_2016}, the information theory is either just mentioned but not really used\,\cite{Sekerka_2015} or totally ignored\,\cite{Swendsen_2019, Olafsen_2019, Luscombe_2021, Pathria_2022}. The situation is also ambiguous in recent research literature.

The contribution of information theory to thermostatistics is two folds that must be well identified. The first one
is linked to the encoding significance of entropy\,\cite{Grunwald_2004} and the relation it gives between energy and the information needed to reproduce the system as it appears to our senses, that is to say make a representation of it.
The second fold, which is no less fundamental, is related to the \textquote{maximum entropy principle} that legitimates an inductive probabilistic inference based on our partial knowledge of the system\,\cite{Jaynes_1957, Jaynes_1968, Shore_1980} to describe its state of equilibrium. It legitimates prior probabilities (the first meaning of probability seen as a degree of belief) as opposed to \textit{a posteriori} probabilities on which a frequentist inference could be done. 

This dual contribution allows for very efficient shortcuts of thought and for resolving inconsistencies in the theory, such as those illustrated by the enigmas mentioned above. This is not new and these advantages are sometime recognized even by those who do not defend the viewpoint of information theory and virulently combat it.
For instance: \myquote{Although information theory is more comprehensive than is statistical mechanics, this very comprehensiveness gives rise to objectionable when it is applied in physics and chemistry} (Denbigh \& Denbigh \cite{Denbigh_1985} p.117).
So, if information theory is not more widely adopted, it is either because its benefits are poorly understood (which we will also try to remedy), or because they are fully understood but rejected for epistemological rather than scientific reasons.
In fact, the stumbling block is that information theory is seen as introducing subjectivity into physics, that is classified as a "hard science" practiced with rigor and objectivity.

In phenomenological thermodynamics, any system is considered as a sort of \textquote{black box} with inputs and outputs in the form of heat and work exchanged with the surroundings. Phenomenological thermodynamics does not care about what is happening at a microscopic level inside the black box. It only deals with these inputs and outputs at the macroscopic scale, which are actually the only measurable quantities.
How can a measurable quantity be subjective? It appears a nonsense. Let us first clarify this point.

The only definition of the entropy $S$ of a system in thermodynamics (Clausius entropy) is in fact that of its variation for a reversible transformation (i.e. sufficiently slow compared to all relaxation processes of the system). Only in this case, it is given by the exact differential $\diff S=T^{-1}\diff Q$ (where $T$ is the temperature and $Q$ the quantity of heat exchanged). But transformations, say from $A$ to $B$, are not reversible is most cases, so that the corresponding variation of entropy cannot be measured. The only way to measure it is to close a thermodynamic cycle by returning to the initial state $A$, this time through a reversible transformation.
The subjectivity of entropy precisely lies here: How to be certain that the cycle is indeed closed? This clearly depends on our knowledge of the initial state. It depend on the informations we have on it:
\myquote{The idea of dissipation of energy depends on the extent of our knowledge} (Maxwell \,\cite{Maxwell_1878}).
\textquote{Objective} means that the quantity under consideration only depends on the object and not on the observer (the subject). In this paper \textquote{subjective} means that the subject (the observer) plays a role. But this role is independent of the person of the subject: two scientists with the same information would reach the same conclusion\,\cite{Jaynes_1968}.
Although this notion of subjectivity thus understood was already present in classical phenomenological thermodynamics, it completely disappeared with the advent of statistical mechanics.

The aim of this paper is, in a first part, to clarify how exactly the subjectivity is brought to thermostatistics via entropy by two means: the encoding of a representation and the probabilistic inference, which are linked to the two features already evoked. Also, the last point will be compared to the alternative frequentist (objectivist) inference, namely the ergodic hypothesis.

The second part addresses the above puzzles and highlights the inconsistencies they raise in relation to the \textquote{objectivist} position. These inconsistencies are all removed with information theory (namely the \textquote{subjectivist} position) in a concise manner.

The last part tries to put the debate at an epistemological level: objective versus subjective conception of entropy. It aims to extricate things, to show where the arbitrariness lies precisely and to answer the question: why is subjective entropy such a disturbing concept?
A particular focus will be done on the filiation of the ideas behind the approach of information theory once applied to thermostatistics.
This filiation corresponds ultimately to a conception about what is science that originates from the Plato's allegory of the cave and develops where modern representationalism (or indirect-realism), empiricism, falsificationism and Bayesianism meet.
That being exposed, everyone can decide whether this conception is natural or worrying, weigh it against the advantages provided by information theory and make a choice.

\tableofcontents

\section{Information }

\subsection{Encoding, information, uncertainty}\label{encoding}

In everyday life, the question \textquote{How much information does this newspaper contain?} is understood in terms of the novelty of the meaning (the substance).
With the advent of communication and computer sciences an alternative signification concerns the minimum quantity of bits that would be needed to transmit, store and reproduce it later (the form).
But, the form is the physical support of the substance and the novelty may lie in the form. Also, an entirely predictable source of \textquote{information} does not require storing data to be reproduced.
So that both acceptations of the term \textquote{quantity of information} are linked.
However, the latter has the advantage to be much more manageable. This was the approach of Shannon: \myquote{[The] semantic
aspects of communication are irrelevant to the engineering problem. The significant aspect is that the actual message is one selected from a set of possible messages. [Transmission and storage devices] must be designed to operate for each possible selection, not just the one which will actually be chosen since this is unknown at the time of design.}\,\cite{Shannon_1948}.
With Shannon, the message becomes a random variable to be lossless encoded and stored.

Consider a source (a thermodynamic system) that sequentially emits a random message $\vx\in \Gamma$ (e.g. adopts a given microstate, $\Gamma$ is the phase space) according to a fixed probability distribution~$p(\vx)$ (the system is at equilibrium).
We plan to perform a lossless recording of the sequence to reproduce it exactly, for instance to study and describe it later.
Whatever the nature of the random events, if the number $\W$ of their possibilities (the cardinality of $\Gamma$ or its \textquote{volume}) is finite, we can establish a one-to-one correspondence table~$\mathcal m$ (a mapping) that assigns to each event $\vx$ an integer $n=\mathcal m(\vx)$ ranging from 1 to $\W$:
\begin{equation}\label{meta1}
	\begin{array}{llll}
	\mathcal m: & \vx\in\Gamma & \longmapsto & n\in\{1,..\W\} 
	\end{array}
\end{equation}
So that recording the system behavior (the source emission) would start by recording the correspondence table (the meta-data), then continue with
recording the sequential outcomes  of the random integer-variable $n$ (the data). This passes through the encoding of the latter, say a binary representation. Thus the question rises: what minimum number $H$ of bits per outcome should be provided for storage or transmission with a given bandwidth?
By quantity of information emitted by the source, we mean this minimum number of bits per outcome for lossless recording.

\subsubsection{Fixed-length encoding}

The central point is to seek the most economical encoding rule\,\cite{Brillouin_1956_book}. A first answer is to plan a fixed-length per outcome (per word). Of course, for the decoder to be able to discriminate the end of a word from the start of the next, this conventional fixed-length must be recorded with the meta-data.
The length must be large enough to store the largest integer $\W$ that is expected to outcome.
Since $\W=2^{\log_2(\W)}$, up to a rounding-error:
\begin{equation}\label{fixed_length}
	H=\log_2(\W)
\end{equation}

The greater the number $\W$ of possibilities, the greater the uncertainty on a given outcome and the greater the minimum length. This last equation allows us to consider $H$,
either as the \textit{a posteriori} average number of bits per outcome, 
or as the \textit{a priori} expected number of bits that should be  scheduled to record upcoming events in case we have absolutely no idea about their actual probability distribution.
In the former case, $H$ is a measure of the quantity of information that has been emitted by the source, whereas in the latter case it is for the observer a measure of the uncertainty about the outcomes. 
These two facets are found in the usual meaning of probability.

\subsubsection{Variable-length encoding}

Equation \ref{fixed_length} is not an optimal solution for storing data, because small numbers, that only require a few bits, take up the same storage space as large numbers.
A variable-length encoding that uses just enough space for each outcome, i.e. $\log_2(m(\vx))$ bits for the outcome~$\vx$, is better.
Of course, this supposes a special encoding, named prefix-code, making it possible for decoding to identify the end of a given word and the beginning of the next. For instance, this can be done by using a delimiter.
This also supposes that the encoding rules are recorded with the rest of the metadata, the size of which, however, will be assumed to be negligible compared to that of the data (which is legitimate for a long sequence of recordings).
With a variable-length encoding, the average number of bits per outcome is:
\begin{equation}\label{variable_length}
	H=\sum_{\vx\in\Gamma} p(\vx)\log_2(\mathcal m(\vx))
\end{equation}
where $p(\vx)$ is the probability of the outcome $\vx$ to which was assigned the integer $\mathcal m(\vx)$ according to the mapping.

Let us first examine the special case where $p(\vx)$ takes the constant value $1/\W$ whatever $\vx$. Eq.\ref{variable_length} gives:
\begin{equation}\label{variable_length2}
	H=\frac{1}{\W}\sum_{\vx\in\Gamma} \log_2(\mathcal m(\vx))
\end{equation}
The smallest values for the series $n=\mathcal m(\vx)$ are obtained by starting from $n_1=1$ 
%(0 is reserved for no event) 
and then by applying the rule $n_i=n_{i-1}+1$. This sequence is that of natural numbers up to $\W$, so one gets:
\begin{equation}\label{variable_length3}
	H=\frac{1}{\W}\sum_{i=1}^{\W} \log_2(i)=\frac{1}{\W}\log_2(\W!)
\end{equation}
For large $\W$, the Stirling formula leads to:
\begin{equation}\label{uniform_distrib}
	H=\log_2(\W) + o(1)
\end{equation}
which is asymptotically the same as Eq.\ref{fixed_length}.
Note that the same result is obtained either if no \textit{a priori} information is known about the outcomes except it is bounded (Eq.\ref{fixed_length}), or if we know in advance that outcomes obey a uniform probability distribution (Eq.\ref{uniform_distrib}).

Variable-length encoding does not make economy of storage-space for a uniform probability distribution. But it remains others.
For non-uniform distributions, it is possible to choose $m(\vx)$ as being an increasing function of improbability $1/p(\vx)$, so that the values that require the most storage space are mapped to the rarest events.
Different rules of assignment can be applied.
For instance, according to the median, split $\Gamma$ into two sub-sets labeled 0 and 1. The first encoding bit for $m(\vx)$ is the label to which subset $\mathbf x$ belongs and the others bits are obtained by subsequent similar recursive dichotomies. This procedure, named Fano encoding\,\cite{Fano_1949}, gives a near optimal encoding length. Shannon\,\cite{Shannon_1948} showed that in no case can the average length per outcome be less than
\begin{equation}\label{Shannon0}
	H = \sum_{\vx\in\Gamma}  p(\vx) \log_2(1/p(\vx))
\end{equation}
Note that the uniform distribution is a special case of this last equation.

\subsubsection{Information encoding and energy}

To a factor $\ln 2$, one recognizes in equations \ref{fixed_length} and \ref{uniform_distrib} the formula for the Boltzmann entropy of  an isolated system (microcanonical), and in equation \ref{Shannon0} that of the Gibbs entropy of a closed system (canonical). In both cases one can write:
\begin{equation}\label{Shannon1}
		S  =  H \ln 2 
\end{equation}
that is called Shannon entropy\,\cite{Price_1982} (in this paper temperature is in Joule, so that entropy is dimensionless).
The Boltzmann-Gibbs entropies are in reality special cases of that of Shannon for which the random events would be the different microstates that a thermodynamic system can adopt.

Here, let us recall that the formula for the Gibb's entropy was obtained from the canonical distribution of energy levels and from 
an identification with certain thermodynamic equalities involving Clausius entropy\,\cite{Lairez_2022a}. The derivation of the Gibbs entropy formula is entirely dependent on thermodynamics.
The Shannon entropy, for its part, is obtained independently of thermodynamics, but with the aim of optimizing the encoding size. Aim which is reminiscent of the idea that Gibbs entropy is maximum at thermodynamic equilibrium.
Equality of the formula and of the idea behind are likely not coincidental. 

The Shannon entropy of the distribution of microstates and Boltzmann-Gibbs entropy are the same quantity.
As the latter is the same as Clausius entropy, the three are one and the same quantity. Hence the connection between energy on one side and information/uncertainty on the other side.

To be more precise about this connection, let us recall some thermodynamics.
Consider a system with internal energy $U$, for any quasistatic process it undergoes, one can write:
\begin{equation}
\Delta U= Q + W,
\end{equation}
where $Q$ is the heat exchanged with the surroundings and $W$ is the work defined as the complementary part of $Q$ in virtue of the conservation law (1st law of thermodynamics).

The 2nd law of thermodynamics is two folds: the first defines the entropy $S$ as a state quantity linked to heat exchanged for a reversible process for which the entropy that is given by its exact differential $\diff S = {\diff{Q}}/{T}$.
Whereas the second is the Clausius inequality that concerns the general irreversible case.
In this paper for the sake of simplicity, we will consider only processes at constant temperature allowing us to more easily integrate $\int T^{-1}\diff Q$. Then, the two folds of the second law become:

\law{2nd law of thermodynamics (\S1)}{\\
There exists a state quantity $S$ which variation for a reversible process is such as $Q=T\Delta S$, where $T$ is the temperature.}

\law{2nd law of thermodynamics (\S2.1)}{\\Clausius inequality: in all cases \begin{equation}\label{cineq1}
	Q\le T\Delta S
	\end{equation}}

\noindent The energy exchanges can be seen as a dissipation ($Q$) of an energy cost ($W$), because generally, heat is unwanted and work is more valued. So that at constant internal energy ($\Delta U=0$), say at constant temperature for a gas,
the Clausius inequality tells us that the energy cost to achieve a process is always greater than  $-T\Delta S$ (i.e. $W\ge -T\Delta S$). 

The twin of Clausius inequality in terms of the quantity of information emitted by the source, or equivalently in terms of the uncertainty about its emission, is obtained with Eq.\ref{Shannon1}, leading to : $W\ge - T\Delta H \ln 2$. The second part of the 2nd law (\S2) can thus be rewritten as

\law{2nd law of thermodynamics (\S2.2)}{\\The energy cost $W$ to vary by $\Delta H$ the uncertainty about the microstate of a system is always such as
	\begin{equation}\label{cineq2}
W\ge - T\Delta H \ln 2
	\end{equation}
	}

\noindent The acquisition of data about the system, via measurement of certain properties (for instance the boundary of the phase space) is a way to reduce the uncertainty about the outcomes. So that the previous equation can be expressed per bit ($\Delta H=-1$) of acquired data:
\begin{equation}\label{Wbit1}
	W_\text{acq/bit} \ge T \ln 2
\end{equation}

\noindent Equations \ref{cineq2} and \ref{Wbit1} are the key equations for linking energy and information. There is nothing more than that.

The two above versions of the 2nd law (\S2.1 and  \S2.2) express exactly the same thing but in different ways.
The first speaks of heat dissipation and entropy, whereas the second prefers to speak of work and uncertainty.
It is thus legitimate to question the real usefulness of the notion of information encoding in thermostatistics.
The first answer is that a link between two fields of knowledge is part of what we call understanding something.
The second answer is that the link between information encoding and energy allows us to express certain ideas in a more concise and consistent way. 
In particular, it provides shortcuts in solving thermostatistic enigmas.
So that if a theory is ultimately an economy of thought\,\cite{Mach_1919, Duhem_2021, Einstein_1934}, this is undoubtedly a progress. The third answer is that the entropy so defined as an uncertainty forms a package with the probabilistic induction that will be discussed in the following section.

\subsubsection{Stability of equilibrium}\label{stability}

The 2nd law of thermodynamics is often expressed for an isolated system to which neither heat nor work is exchanged with the surroundings ($Q=W=0$). So that Eq.\,\ref{cineq1} becomes 
\begin{equation}
	T\Delta S\ge0
\end{equation}
This leads to another version of the second law:

\law{2nd law of thermodynamics (\S2.3)}{\\The entropy of an isolated system cannot spontaneously decrease.}

\noindent Or in terms of information:

\law{2nd law of thermodynamics (\S2.4)}{\\The uncertainty (the quantity of information) about the microstate of an isolated system cannot spontaneously decrease.}

Classical and phenomenological thermodynamics is traditionally only concerned with equilibrium \textit{stricto sensu}: \myquote{a system is in an equilibrium state if its properties are consistently described by thermodynamic theory} (H.B. Callen\,\cite{Callen_1985} p.\,15). 
In classical thermodynamics the equilibrium is by definition a stationary stable state and \textquote{equilibrium state} is a pleonasm: no states other than those at equilibrium are defined (by state quantities).

With the Boltzmann-Gibbs entropy, probabilities come into play for the description of the thermodynamic equilibrium. 
Therefore, an isolated system can now fluctuate and deviate slightly from equilibrium. As this notion was absent in phenomenological thermodynamics, we are now faced with this problem: How can the equilibrium be stable? What is the restoring force of the system when it deviates from equilibrium? To ensure the stability of equilibrium, if we do not want to postulate it, we need an additional ingredient under the form of an alternative postulate or a definition of the nature of equilibrium. This definition can be the following: 

\law{Definition of equilibrium (v1)}{\\The equilibrium of an isolated system is the state of maximum entropy.}

\noindent Or equivalently

\law{Definition of equilibrium (v2)}{\\The equilibrium of an isolated system is the state of maximum uncertainty about its microstate.}

\noindent With these definitions, the restoring force comes from the 2nd law: due to fluctuations, the system may deviate from equilibrium but will return to it spontaneously.

\subsection{Inductive probabilistic inference}\label{induction}

The first programme of statistical mechanics is to calculate certain observable macroscopic quantities of a thermodynamic system at equilibrium, from the average of certain random variables which are relevant at a microscopic level.
For instance, calculating the temperature from the average kinetic energy of particles.
These averages are computed over a probability distribution.

The central point is thus to determine which random variable to consider and what probability distribution it is supposed to obey. That is to say make a statistical inference.

\subsubsection{Subjective versus objective probabilities}

Two different types of statistical inferences are traditionally distinguished: \textquote{probabilistic inference} and \textquote{frequentist inference}, which depends on how probabilities are defined.
\begin{enumerate}
	\item subjective (or prior) probabilities are reasonable expectations or degrees of belief that one thing or another will happen. They are subjective in that they depend on our knowledge of the system.
	\item objective relative frequencies of occurrence (over an ensemble) of one thing or another that actually happened (or \textit{a posteriori} probabilities). They are supposed to be a tangible property of the system.
\end{enumerate}

Subjective probabilities are general and can always apply, so they are \textit{de facto} the most common on which to base a  decision. But their arbitrary nature poses a problem without a rational criterion to assign a value to them. They appear illegitimate and may turn out to be false \textit{a posteriori}.
In contrast, frequencies are reliable provided the corresponding measurement has been carried out. But this is often impossible or at least not possible before making a decision. Their use is conditioned on the existence of an ensemble, or at least on the hypothesis of its existence provided it can be done in a consistent manner.

Consider the game of die.
The die is cubic and offer six possible outcomes.
Also from a symmetry argument there is no reason to believe that one is more likely than another.
Prior to any toss of die, we can reasonably assign to outcomes a uniform discrete probability distribution lying from 1 to 6. This reasoning, which accords with common sense\,\cite{Jaynes_1988}, is called \textquote{Laplace's principle of insufficient reason} (or \textquote{principle of indifference}\,\cite{Keynes_1921}). It is a typical exampe of probabilistic inference.

An interesting point is that the most reasonable decision for a bet would be exactly the same if we know in advance that the die is loaded (we know in advance that the distribution is not uniform), but do not know which number is favored.
The first assignment of probabilities is ultimately based on a criterion that seems much more arbitrary (\textquote{there is no reason to believe otherwise}) than waiting for a few tosses of die and estimating the frequency distribution from the sampling of outcomes (make a frequentist inference). The decision is based on prior probabilities which do not seem legitimate, but are the only ones available.

In the problem of information processing, faced with an unknown source that we want to record, we must begin by using one or other of the encoding rules 
%(take a decision we expect as much as rational as it can be) 
and then eventually use an adaptive procedure to reconsider (to update) the encoding according to the observations. 
For a lossless recording, the best choice to begin is a fixed-length encoding (with an overestimated number of possible outcomes if it is unknown). This choice maximizes the compatibility of the encoding procedure with future incoming data.
In section \ref{encoding} we saw that this choice amounts to assigning a uniform prior distribution for the outcomes and is therefore in agreement with the principle of insufficient reason.
We know in advance that this prior distribution is certainly not the true one, but this does not prevent it from being chosen rationally. This choice is the best we can make, any other would be judged irrational.

The problem of assigning a prior distribution is ultimately reduced to the search for an optimal compromise between two contradictory goals: 1)~avoid any loss of information; 2)~avoid unnecessary volume of storage.
This problem of optimization is formalized and generalized with another aspect of the Shannon's information theory that essentially aims to optimize the use of our prior knowledge of the source for the statistical inference.

\subsubsection{Maximum entropy probabilistic inference}\label{maxentinf}

Suppose we are dealing with a source that emits outcomes $i$ about which we have only a partial knowledge of the true probability distribution $p_i$, so that different distributions can potentially satisfy these constraints. 
To start recording the source, we need to assign a prior distribution that \myquote{agrees with what is known, but expresses a 'maximum uncertainty' with respect to all other matters} (Jaynes\,\cite{Jaynes_1968}) and thus leaves a maximum chance of compatibility with subsequent data.
The problem is reduced to maximizing the uncertainty, which therefore remains to be measured.

In addition to the definition of the entropy as the optimum length to encode the outcomes of a source, Shannon\,\cite{Shannon_1948} introduces another idea.
He asks \myquote{Can we find a measure [...] of how uncertain we are of the outcome?}, and continues \myquote{If there is such a measure, say $H(p_1, p_2, ...)$, it is reasonable to require of it the following properties:} 1)~being continuous in $p_i$; 2)~being increasing in $\mathcal W=1/p_i$ for a uniform distribution; 3)~being additive over different independent sources of uncertainty. Then, Shannon demonstrated that the only function (the only measure of uncertainty) that fulfills these requirements is to a factor $\sum p_i \ln(1/p_i)$, i.e. the Shannon entropy. Hence the theorem: 

\law{Maximum-entropy theorem}{the best prior distribution $p(x$) that maximizes the uncertainty on $x$ while being consistent with our knowledge, is the one that maximizes Shannon entropy.}

The validity of this theorem is ultimately determined by what is supposed to be required for a measure of uncertainty. These requirements play the role of starting axioms to its demonstration. As natural as they seem, %(to be convinced read \cite{Cox_1946, Jaynes_1988}), 
some may find them arbitrary, wondering why not prefer others, leading to another function to be maximized. 
Shore and Johnson\,\cite{Shore_1980} start with a completely different requirement, a consistency axiom that can be stated like this: \myquote{if the problem of assigning a prior distribution can be solved in more than one way of taking the same information into account (for instance in different coordinate systems), the results should be consistent}. On which everyone should agree.
Then, Shore and Johnson\,\cite{Shore_1980} prove that the only procedure satisfying this requirement of uniqueness is that of maximizing Shannon entropy. Given a random variable $\mathbf x$, the maximum entropy theorem provides a legitimate method to determine the prior probability distribution $p(\vx)$.

However, to complete the programme of statistical mechanics, it remains the first point risen in the beginning of this section: which random variable $\mathbf x$ to consider.
For instance, imagine a random variable $x\in[0,\pi]$ with a uniform distribution, $\sin(x)$ is also a random variable which distribution is not uniform but has a maximum for $x=\pi/2$. Thus, using the maximum-entropy criterion directly on $x$ or on $\sin(x)$ lead to different distributions \textit{in fine} for $x$.
However, \myquote{among all these distributions there is one particular one, corresponding to the absolute maximum of entropy, which represents absolutely stable equilibrium.} (Planck\,\cite{Planck_1914} p.32).
Jaynes\,\cite{Jaynes_1973} outlines a crucial point. The assumption of uniqueness of equilibrium actually automatically brings to our knowledge others crucial informations: the solution is not supposed to depend on the orientation of the observer (invariance under rotation), nor on its position (invariance under translation), nor on the scale it is considered (invariance under scaling). 
Among all the possible variables describing a system, considering only those whose distributions are invariant in form under similarities avoids inconsistent results (lead to the same result).
Hence the definition of equilibrium:

\law{Definition of equilibrium (v3)}{\\The equilibrium of a system is the only state that maximizes the uncertainty on variables whose distributions are similarity-invariant in form.}

\noindent This definition plus the maximum-entropy theorem, can be expressed all-in-one under the form of the so called  maximum-entropy principle that is actually also an alternate definition of the equilibrium:

\law{Definition of equilibrium (v4)}{\\The equilibrium of a system is the only state that maximizes Shannon entropy of variables whose distributions are similarity-invariant in form.}

Two classical examples of similarity-invariant distributions mentioned in the above definitions are those of Boltzmann and Gibbs:
\begin{itemize}
	
\item Boltzmann\,\cite{Boltzmann_Lectures} considers the phase-space of one single particle ($\Gamma_1\subset\mathbb{R}^{6}$). For $N$ particles, provided that their phase spaces are identical, the probability distribution $p(\vx)$ he considers is that of finding one particle in a given elementary \textquote{volume}  $\vx$ ($p(\vx)$ is the particle density).  The H-function he defines is the Shannons's entropy of $p(\vx)$.

\item Gibbs\,\cite{Gibbs_1902}, considers directly the phase space of $N$ particles ($\Gamma\subset\mathbb{R}^{6N}$).
$p(\vx)$ is the probability for the system to be in microstate $\vx$. The corresponding Shannon's entropy is the same as that defined by Gibbs who showed that it is also equal to the Clausius entropy of the system.

\end{itemize}

Maximizing the entropy of either distribution yields two consistent descriptions of equilibrium.
In practice, the maximization procedure is a variational calculus  taking into account the constraints imposed by our knowledge about the system\,\cite{Shannon_1948}. For instance, suppose that the only thing we know about $p(\vx)$ is that it has a finite support (the minimum required for a discrete distribution to be properly normalized), then the best distribution is uniform (this is the microcanonical distribution for an isolated system). If $p(\vx)$ is only known to have a positive support allowing $\vx$ to have a finite average value, then the best distribution is the exponential decay (this is the canonical distribution of energy levels for a closed thermalized system).

In these last two definitions of equilibrium (v3 and v4), the statement of uniqueness should not be controversial.
The core of the controversy lies in the rest. 
One may consider that there is absolutely no reason why the system would actually maximize the uncertainty we, observers, have about its microstate.
But making another inference would be neither optimal nor reasonable.
Moreover, the similitude between the first definitions of equilibrium (v1 and v2) that were given in section \ref{stability} and these last two is noticeable.
Whereas the first definitions were based on thermodynamical considerations and from the need for the theory to define the equilibrium as stable, the second are more general (not only concerned with microstates) and emerge from a reasoning totally free from thermodynamics and free from any considerations regarding stability.
This reasoning is an inductive probabilistic inference, that will be called \textquote{subjective} in contrast with the alternative, called \textquote{frequentist}, that is believed to be more objective.

\subsubsection{Alternative \textquote{frequentist} inference}\label{ergodicity}

Statistical mechanics did not wait for information theory to infer distributions at equilibrium. Alternatives approaches focus on the distribution of microstates. In addition,
the problem lies uniquely in deciding what is the distribution for an isolated system. Because, this being determined, the distribution for any closed subpart can be deduced\,\cite{Lairez_2022a}.
These alternative approaches are essentially of two kinds.

The first is based on the already mentioned principle of insufficient reason, 
which is renamed for the circumstance \textquote{fundamental postulate of statistical mechanics}\,\cite{Balian_1991}.
In fact, it is nothing other than a less formalized and general expression of the maximum entropy principle\,\cite{Uffink_1995}.

We will only focus on the second alternative approach, that of \textquote{frequentists}, which intends to adopt an objective point of view. To compensate for the lack of knowledge of the system, the idea is to make a strong hypothesis, that of \textquote{ergodicity}. 
This hypothesis essentially aims to fulfill the prerequisite for the definition of probabilities as relative frequencies: the existence of an ensemble on which to calculate them.

Consider an isolated thermodynamic system, its phase space $\Gamma$ is the set (ensemble) of all possible microstates $\vx$ of probability $p(\vx)$ under which copies of the system can be found. Take one of this copy. It is dynamical, that is to say it continuously undergoes over time a transformation $F: \Gamma\mapsto\Gamma$, allowing to define, from the initial condition $\vx_0$, a trajectory (an orbit) as the set of points in the phase space $\mathcal T(t)=\{\vx_0, F(\vx_0),F^2(\vx_0), ...F^t(\vx_0)\}$. 
The ergodic hypothesis is that this trajectory will finally pass recurrently through all points of the phase space at the frequency $p$, that is therefore supposed to remain unchanged over time (the transformation $F$ preserves the volume of the phase space).

For our concern to determine which probability distribution is that of microstates, one consequence of the ergodic hypothesis is that \textquote{volume preserving} transformation means for any point $\vx$ of the trajectory
\begin{equation}\label{volpres}
	p(\vx)=p(F^{-1 }(\vx))
\end{equation}
That is, the probability of a given consequent microstate is that of its antecedent. 
%The trajectory is seen as a chain of causation.
The probability distribution is therefore uniform and unchanged throughout the trajectory which will ultimately cover the entire phase space. We thus obtain the microcanonical distribution we were looking for.

For comparison with the maximum entropy approach, it is interesting to express the hypothesis of ergodicity in the same form as previously, i.e in the form of a definition of equilibrium:

\law{Definition of equilibrium (v5)}{\\The equilibrium of a system is the only state where the distribution of microstates is the same over its time-transformation as over an ensemble of its copies.}

\noindent It immediately appears that this definition, contrary to the previous ones, does not contain any warranty of stability of the equilibrium. There is no restoring force for the equilibrium, which so defined is not an attractor for the system. This point is one of the source of the inconsistencies that will be discussed in the following.

At the basis of the hypothesis of ergodicity is the fact that deterministic Hamiltonian systems, 
 (according  to the Liouville's theorem) are volume preserving and thus ergodic.
  In this context, the trajectory of the system in the phase space is a chain of causality. 
So that Eq.\ref{volpres} is more particularly interpreted as: the probability of a given consequent microstate is that of its \textquote{cause}.

Still for the sake of comparison, the detailed logical steps of the inductive reasoning for the ergodic hypothesis, which makes it natural to us, are the following: 
1)~a set of atoms (a thermodynamic system) is analogous to a set of colliding rigid spheres (in the time of Maxwell and Boltzmann, as the notion of atom itself, this was far from obvious);
2)~usually, a set of colliding rigid spheres comes under classical mechanics;
3)~usually, the equations of motion of mechanics alone determine the future state of classic mechanical systems (this is generally the meaning of the word \textquote{deterministic}, but there exist exceptions of non-deterministic classic mechanical systems, e.g. the \textquote{Norton's dome}\,\cite{Norton_2003b}).
It follows that: 4)~given a microstate $\vx$ adopted by the system, the equations of motion of classical mechanics alone determine the future microstate $F(\vx)$ (Eq.\ref{volpres}).

In short, there is no definitive proof for the system to be ergodic as measurements is not possible, but we consider this very likely. The reasoning is ultimately not that far from a probabilistic inference, but much less explicit on this point than that of maximum-entropy.

%It appears that this reasoning is also based on inductive inferences: generalization (points 1 and 2) and analogy (point 3). In fact, in natural sciences, inductive reasoning is inevitable as it will be recalled in section \ref{subphy}.

\section{Enigmas}

Information theory introduces subjectivity in thermostatistics by two different and related manners: 1)~the encoding of a representation of the system and its link with energy; 2)~a probabilistic inference. In what follows, these two features are used to resolves the inconsistencies raised by the thermostatistic enigmas quoted in the introduction.

Here, these enigmas are classified into two categories, paradoxes and demons, which have not the same level of importance.
Paradoxes raise inconsistencies that cannot be removed by classical statistical mechanics without information theory.
As for demons and the devices they drive, they  do not actually introduce inconsistencies, in so far as they are physical systems that obey to the second law of thermodynamics, as do all others.
But they can rather be considered as the first evidences of the link between information and energy that were given by Maxwell, Zermello and Szilard well before Shannon.

\subsection{Gibbs paradoxes}

\subsubsection{The problems}

\begin{figure}[!htbp]
	\begin{center}
		\includegraphics[width=1\linewidth]{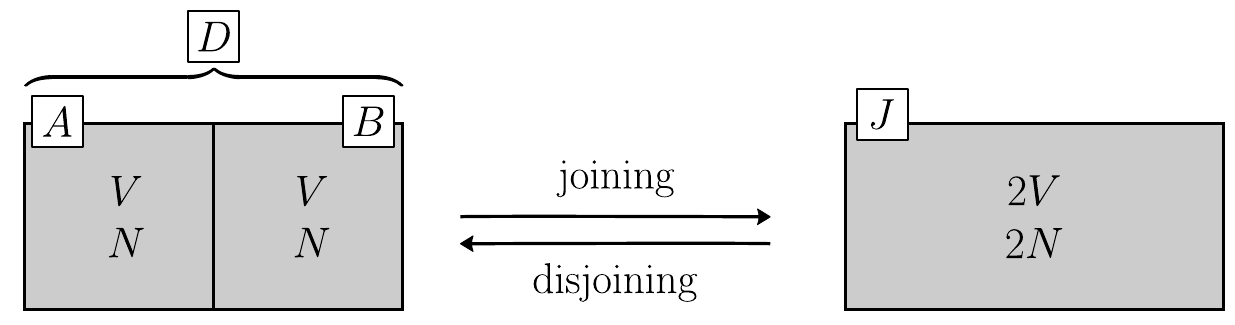}
		\caption{Usual image of the joining-disjoining thermodynamic cycle of two volumes of gas: removing the partition joins the two volumes and putting it back restores the system to its initial state with no work nor heat exchanged with the surroundings.}
		\label{join1}
	\end{center}
	
\end{figure}

Consider a system $D$ (for disjoined) made of two adiabatic containers, denoted $A$ and $B$, of same volume $V$, each containing the same quantity $N$ of an ideal gas (of the same species or not) with the same degree of freedom and at the same temperature~$T$ (see Fig.\ref{join1}). So that, the entropy of the two sub-systems are equal $S=S_A=S_B$.
From the additivity of entropy:
\begin{equation}
	S_D=2S
\end{equation}
Joining these two volumes by removing the partition between them results in another system $J$ (for joined). The process is never accompanied by any observable thermodynamic effect: neither heat nor work is exchanged with the surroundings. 
To determine whether or not this is accompanied by an increase of Clausius entropy, we must go back to the disjoined state (close the cycle) by a reversible process and measure heat exchange.
Here, two paradoxes arise and continue to be debated for 150 years (for a review of the debate see e.g. the papers in \cite{Gibbsparadox2009, Gibbsparadox2018}).

 If the two gases are identical, according to thermodynamic phenomenology replacing the partition between the two containers allows the system to return to its initial state. This is done without heat exchange leading to the conclusion that the two states have the same Clausius entropy.
\begin{equation}\label{mix_clausius0}
\Delta S=S_J-S_D=0
\end{equation}
However, if initially the two gases differ, removing the partition mixes them and just putting it back is insufficient to separate them again. The separation can be done by two isothermal compressions against two pistons equipped with different semi-permeable membranes\,\cite{Planck_1903}.
The first piston is only able to compress one species (from $2V$ to $V$) and the second only the other species (by the same ratio). The total work is thus equal to $W=-Q=T\Delta S=T\times2N\ln 2$. Leading to
\begin{equation}\label{mix_clausius1}
	\Delta S =S_J-S_D= 2N\ln 2
\end{equation}
In Eq.\ref{mix_clausius0} and \ref{mix_clausius1}, $\Delta S$ is called entropy of mixing.

The first paradox mentioned by Gibbs\,\cite{Gibbs1874} is that $\Delta S$ is a bivalued discontinuous step function of the dissimilarity of the gases, whereas it can legitimately be expected to be continuous like other property-variations of the system (density, refractive index...).

The second paradox comes from statistical mechanics.
The Boltzmann entropy of the disjoined state is $S_D=2S=2N\ln V$, whereas it is $S_J=2N\ln 2V$ for the joined state. Leading to the difference $\Delta S=2N\ln 2$. This, whatever the gas species, identical or not. If statistical mechanics solves the paradox of discontinuity, it raises another: Why do the Boltzmann and Clausius entropies differ when they should be the same?

\subsubsection{Usual solutions}

Concerning the first paradox, the consensus is that the discontinuity is not problematic. In fact, the dissimilarity of two species of atoms is discontinuous, thus that of entropy is not a problem. In term of the classification of Quine\,\cite{Quine1976}, the discontinuity is treated in the literature as a veridical paradox (the two premises are correct but not inconsistent). 

As for the second paradox, it is most of the time treated as a falsidical paradox (at least one premise is wrong). As phenomenology is the final arbiter, the calculation of the cardinality of the phase spaces must be reconsidered (corrected). 

Justifications for this correction are mainly of two kinds.
Denote $\V_D=\V^2$ and $\V_J$ the cardinality of the phase spaces for systems $D$ and $J$, respectively, in the case where all the particles are different and clearly identified with a label, such as a serial number. 
%The cardinality of the disjoined state is: $V_D=\V_A\V_B$
Also, denote $\W_D=\W^2$ and $\W_J$ the corresponding cardinality in the case where all particles are identical.

The first approach to justify a correction is based on the notion of indistinguishability of particles that comes from quantum mechanics\,\cite{Huang_1987}. For $N$ indistinguishable particles, their $N!$ permutations give the same microstate which must therefore be counted only once leading to $\W=\V/N!$. This is known as the correct Boltzmann counting, leading to:
\begin{equation}\label{quant}
	\W_D=\frac{\V_D}{N!N!}, \quad 	\W_J=\frac{\V_J}{(2N)!}
\end{equation}

The second approach remains within the framework of classical mechanics where particles (even identical) are always distinguishable. In the sense that they always have distinct trajectories, allowing them to be (in principle) traceable and thus identified at any time.
When partitioning the system into two compartments, particles can be combined in $(2N!)/N!N!$ different manners into the two separate compartments\,\cite{Cheng_2009, Versteegh2011, Frenkel_2014}. It follows that the number of possible results for the disjoined state is increased by this multiplicative factor, leading to:
\begin{equation}\label{class}
	\W_D=\V_D \frac{(2N)!}{N!N!}, \quad 	\W_J=\V_J
\end{equation}

It follows that with the two approaches, the entropy of mixing two identical gases is the same:
\begin{equation}
	\Delta S = \ln\left(\frac{\W_J}{\W_D}\right)=\ln\left(\frac{\V_J}{\V_D}\frac{N!N!}{(2N)!}\right)
\end{equation}
but for different reasons\,\cite{Dieks_2014}. In both cases, the second Gibbs paradox is claimed to be solved, because by using an approximation of the Stirling formula one gets $\ln(N!N!/(2N)!)=2N\ln(N)-2N\ln(2N)=-2N\ln 2$. So that the excess of entropy (Eq.\ref{mix_clausius1}) obtained with the Boltzmann equation is corrected.

It is important to outline that Eq.\ref{class} for $\W_D$ counts the number of all possible disjoined microstates. That is to say, $\W_D$ is the cardinality of the phase space viewed as an ensemble of different possibilities including the different possible combinations in the repartition of particles. 
But once the partition is in place, a given disjoined microstate thus obtained will never by itself have a dynamic trajectory allowing it to reach another repartition (the repartition is frozen). 
In other words, the dynamics of a disjoined state cannot allows all the possibilities accounted for by Eq.\ref{class} to be explored. 
The disjoined state is no longer ergodic (this is noted in \cite{Peters_2013}).
It follows that Eq.\ref{class} is implicitly valid if the corresponding entropy is an uncertainty about the actual state of the disjoined system. Probabilities are prior-probabilities and not frequencies. In a classical mechanics framework, the above solution automatically places us implicitly in a \textquote{subjectivist} rather than in a \textquote{frequentist} position. That is to say, there is no solution of the 2nd Gibbs paradox in the framework of classical statistical mechanics and frequentist (ergodic) inference. The solution is necessarily quantum or based on a probabilistic inference. It is up to the reader to decide which one is more consistent and natural.
This often goes unnoticed.

The \textquote{subjectivist} position to solve the 2nd Gibbs paradox is explicit in some papers, which are nevertheless largely in a minority (see e.g. \cite{van_Kampen_1984, Jaynes1992, Tseng_2002, Dieks_2013}). But even in these latter papers, the paradox of discontinuity is either eluded of treated as veridical. 

\subsubsection{Yet another solution}

The aim of this section is to show how from the information point of view,
 the dissimilarity of two gaseous contents is not bivalued but gradual and the Shannon's entropy too, that is, as close as possible to a continuous function with atomistic matter. So that the paradox of discontinuity is actually falsidical. In doing so, the 2nd paradox is also solved by considering it to be veridical.
 
\begin{figure}[!htbp]
	\begin{center}
		\includegraphics[width=1\linewidth]{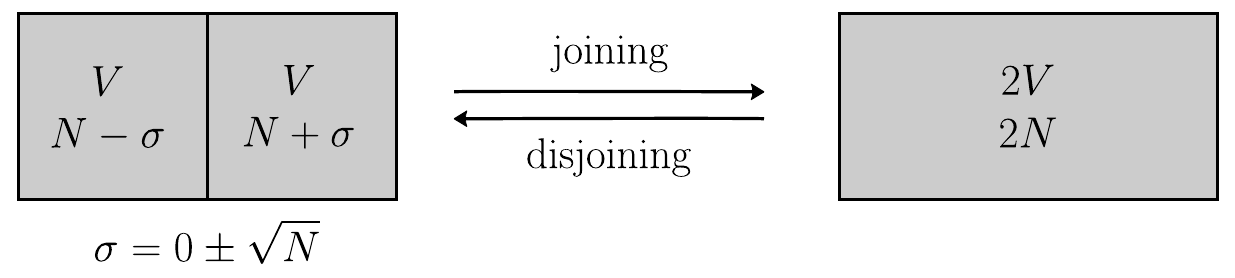}
		\caption{Correct image of a joining-disjoining thermodynamic cycle supposed to be repeatable (to be compared to Fig.\ref{join1}): putting back the partition leaves each compartment with the same number of particles up to the standard deviation $\sqrt N$.}
		\label{join2}
	\end{center}
	
\end{figure}

Let us start by observing that thermodynamics considers cycles performed repeatedly and reproducibly. Therefore, if the two gases are identical, the representation of a joining/disjoining cycle does not care about:
\begin{enumerate}
	\item the exact number of particles in each compartment
	up to the standard deviation $\sqrt N$ of the binomial distribution; 
	\item the traceability of particles (this information is only relevant if the two gases initially differ).
\end{enumerate} 
The correct image for the joining-disjoining cycle is given by Fig.\ref{join2} (instead of Fig.\ref{join1}). It is to this very cycle that thermodynamics refers in Eq.\ref{mix_clausius0}.
The calculation of $\Delta S$ in term of probabilities has thus to be done by accounting for these two useless pieces of information.
Accounting for the latter is very common and leads to the correct Boltzmann counting ($-\ln N!$ term), accounting for the former requires in addition the use of the exact Stirling formula $\ln(N!)  = N \ln(N/e) + \ln(\sqrt{2\pi N}) + o(1)$ (rather than the usual approximation that consists in the first term only). It can be found in \cite{Lairez_Stirling}.
Here, a different derivation is proposed that avoids the term $-\ln(N!)$ and allows us in doing so to solve the paradox of discontinuity.

\begin{figure}[!htbp]
	\begin{center}
		\includegraphics[width=1\linewidth]{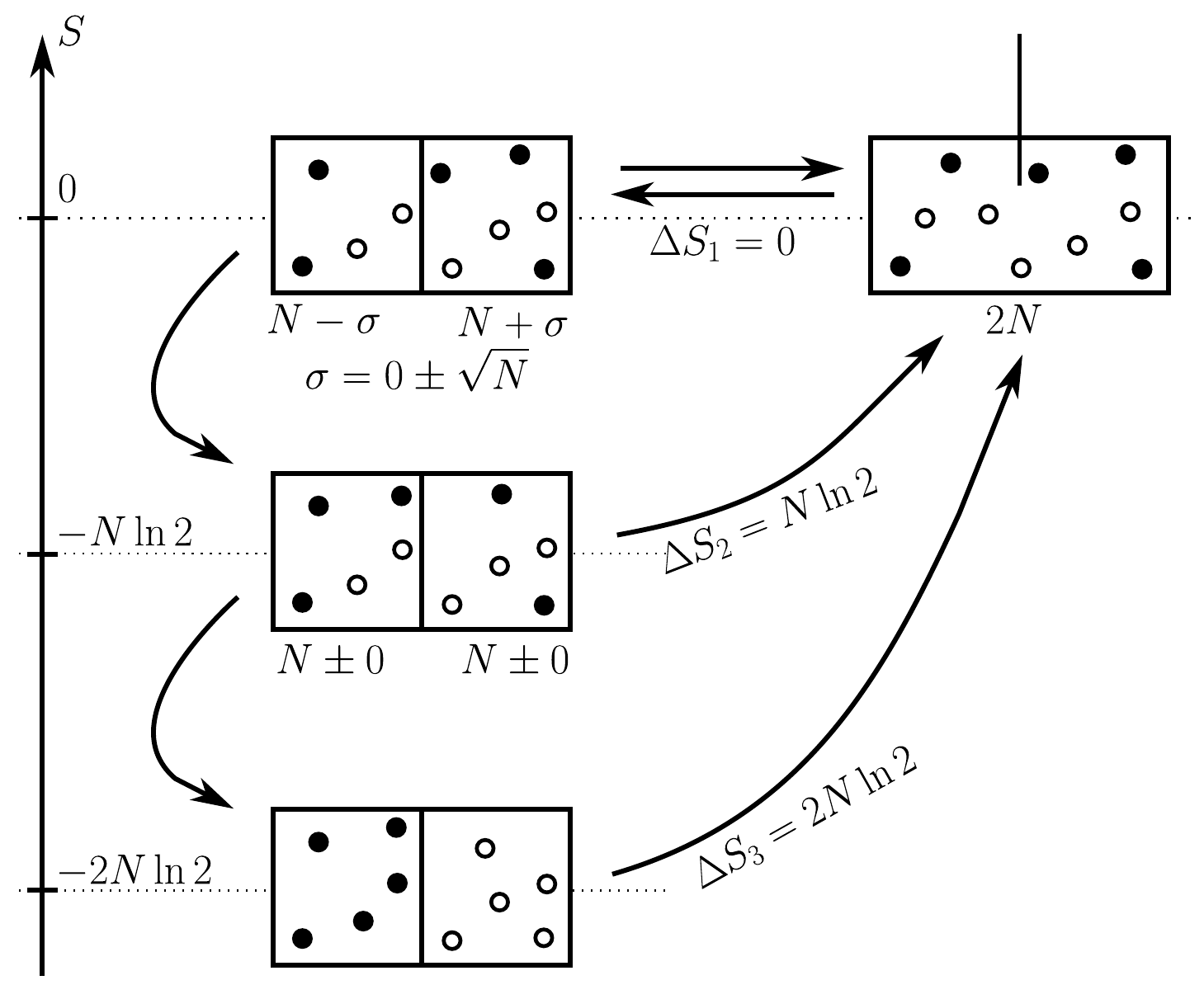}
		\caption{Mixing-unmixing cycle of a gas made of two species. The cycle depends on the information we had and do not want to lose. 
		Top: information about exact number of particles per compartment and traceability is lost at the end of the first cycle, therefore further cycles leave it unchanged ($\Delta S_1=0$).
Middle: retrieving exactly the same number of particles in each compartment has a minimal energy cost of $N\ln 2$ ($\Delta S_2=N\ln 2$). Bottom: preserving particles traceability has an additional minimal energy cost also equal to $N\ln 2$ ($\Delta S_3=2N\ln 2$)}
		\label{mixcycle}
	\end{center}
\end{figure}

Consider a first kind of cycle (see Fig.\ref{mixcycle} top) that consists in just moving the partition between the two compartments, denoted $A$ and $B$. Let $t_0$ be the time just before the first cycle starts. After the first cycle is ended, all pieces of information about the exact contents of $A$ and $B$ at time $t_0$ is lost.
	At the end of each cycle in the disjoined state, the number of particles per compartment is always $N\pm\sqrt{N}$.
	Also after $t_0$, any information about the traceability of particles is lost. 
	So that the uncertainty concerning these two features, exact number of particles and traceability, is unchanged by further cycles, and so the Shannon entropy. Therefore by considering these further cycles, the Shannon entropy of mixing is zero, just like Clausius entropy:
	\begin{equation}\label{Smix0}
		\Delta S_{1}= 0
	\end{equation}

Imagine that we know for certain that initially the two compartments had exactly the same number of particles $N\pm0$ and that we want to retrieve
this information when restoring the disjoined state (see Fig.\ref{mixcycle} middle).
The procedure for the gas partition can be the following:
\begin{enumerate}
\item put all particles in a separate box;
\item partition the empty volume $2V$ into $A$ and $B$;
\item take iteratively one pair $(a,b)$ of particles, put one particle (either $a$ or $b$) in compartment $A$ and the other in $B$.
\item after $N$ iterations, the two compartments have exactly the same number of particles.
\end{enumerate}
There are four possibilities to arrange $a$ and $b$ in two boxes:  $\{.|ab , a|b, b|a, ab|.\}$, and only $a|b$ or $b|a$ are convenient. 
Therefore each iteration divides the number of possibilities by 2 and gives us 1 bit of information. The $N$ iterations of the overall procedure and Eq.\ref{Wbit1} provide the corresponding entropy of mixing:
\begin{equation}\label{Smix1}
	\Delta S_{2} = N \ln 2
\end{equation}
Note that the procedure can be stopped at any iteration, if we are satisfied by the uncertainty on $N$ would lead a random repartition of the rest of particles. So that depending on our wish, the entropy of mixing can take any value from 0 to $N \ln 2$ by step of $\ln 2$.

Imagine that we know for certain that at $t_0$, compartment $A$ was filled with isotope $a$ and compartment $B$ with isotope $b$ (see Fig.\ref{mixcycle} bottom). So that, we are not satisfied by the previous procedure and want to restore exactly the original state. In other words, we want to preserve the traceability.
To achieve this, among the two possibilities $\{a|b, b|a\}$ in the previous procedure, we must choose $a|b$. Here again, at each iteration the number of possibilities is divided by 2 and gives us 1 additional bit of information. So that at the end, compared to the previous state the entropy has decreased by an additional amount $N\ln 2$. Finally, if we consider traceability as crucial the Shannon entropy of mixing is:
\begin{equation}\label{Smix2}
	\Delta S_{3} = 2N \ln 2
\end{equation}
Here again, the procedure can be stopped at any iteration according to which degree of  impurities is acceptable. 

Depending on our knowledge about the original state or depending on what we consider as being important about it, the mixing-unmixing cycle differs and the Shannon entropy of mixing too. The latter can take gradually any value from 0 to $2N \ln 2$ by step of $\ln 2$. The two Gibbs paradoxes are solved.

\subsection{Paradoxes related to H-function}

%Poincaré-Zermelo recurrence paradox}

Boltzmann was the first\,\cite{Boltzmann_Lectures} to write a quantity defined at any time $t$, named H-function, that takes the form of an entropy.
For that, he considers the phase-space of one single particle ($\Gamma_1\subset\mathbb{R}^{6}$) and the probability $p(\vx)$ to find a particles at $\vx\in\Gamma_1$ at time $t$ (with a time scale supposed to be discretized like the phase space).
The H-function is defined as:
\begin{equation}\label{defH}
	H(t) = \sum_{\vx\in\Gamma_1} p(\vx, t)\ln( 1/p(\vx, t))
\end{equation}
which is nothing other that the Shannon's entropy of the particle density $p$.

Starting from the idea of the Maxwell's kinetics theory of gases that the motion of colliding rigid particles is governed by the equations of classical mechanics, but also that their large number allows a statistical treatment, Boltzmann has obtained an integro-differential equation (named Boltzmann transport equation) for the time variation of the density $p$ of a dilute gas (for a reference book see \cite{Cercignani_1988}). Boltzmann's equation allowed him to prove that for a bounded phase-space (an isolated system):
\begin{equation}\label{Htheo}
	\frac{\diff H}{\diff t}\ge 0
\end{equation}
with the equality corresponding to the equilibrium state defined as stationary.
Eq.\,\ref{Htheo} is known as the H-theorem.
\myquote{Its proof is clever and beautiful, although not perfectly rigorous.}
(Villani\,\cite{Villani_2008}). In fact, the proof in the general case is still in progress.
But for physicists, the H-theorem is quite natural and can be viewed as another expression of the 2nd law of thermodynamics\,\cite{Weaver_2021} (concerning another distribution than that of microstates) plus a definition of equilibrium that warrants its stability. 
The 2nd law states that the entropy of an isolated system cannot spontaneously decrease.
So that, even if classical thermodynamic (that of Clausius) says nothing about what exactly happens during a spontaneous process, but only deals with the entropy before and after (the time variable is not present in classical thermodynamic equations), it is legitimate to say that entropy increases with time during a spontaneous process.
For instance, consider a gas which is initially confined in a small box inside a larger room. Opening the box allows the gas to expand freely over an increasing volume ($\diff H/\diff t>0$) until it uniformly occupies the entire room ($\diff H/\diff t=0$) in a stationary state. In doing so, its entropy continuously increases.

The Boltzmann transport equation and H-theorem constitute the first attempt to demonstrate the macroscopic 2nd law of thermodynamics from what happens at the microscopic scale.
Against this attempt, two objections have been raised: the reversibility paradox and the recurrence paradox.

\subsubsection{Loschmidt's reversibility paradox}\label{reverse}

This paradox  was originally stated\,\cite{Darrigol_2021} under the form of a thought experiment.
Consider the free expansion of a gas enclosed initially at time $t_0$ in a box placed in a larger room.
Once the system is at equilibrium, after a certain time $\tau$, imagine that the direction of the velocity of each particle is reversed, without changing its magnitude. 
The operation does not change the macroscopic properties of the gas, such as temperature or volume. So that neither heat nor work are provided to the system.
But, once this has been done, the gas particles goes backward through the same sequence of collisions than the previous one. So that at time $2\tau$ its original microstate is restored. The gas is returned inside the box in contradiction with the 2nd law of thermodynamics and with the observation that this never happens.

The problem can be viewed in two different manners. First, to decide whether or not the process violates the 2nd law, we must wonder how the operation of reversing the velocities of particles is possible and whether it can be done without energy expenditure (certainly not if the operation is physically performed with a device that obeys to the 2nd law). This problem, thus posed in term of an operating \textquote{demon}, will be discussed in section  \ref{Loschmidt_demon}.

The other viewpoint is that this paradox basically raises the question of how from time-symmetrical equations of motion (those of mechanics) it is possible to obtain time-asymmetrical results.
The consensual answer\,\cite{Uffink_2006, Weaver_2021, Weaver_2022} is that, within the ingredients that permit to write the Boltzmann transport equation, the time-asymmetry is already presents under the form of the \textquote{hypothesis of molecular chaos}:
the velocities of two particles before their collision are fully uncorrelated but of course fully correlated and determined by mechanics after the collision.
Fundamentally, the Boltzmann transport equation (and thus the H-theorem) is obtained by moving the time asymmetry of the 2nd law of thermodynamics from the macroscopic to the microscopic scale. The 2nd law is phenomenological and comes from an inductive reasoning which basically is a generalization of observations. By moving at the microscopic scale, it becomes a postulate or a hypothesis allowing to build a deductive reasoning.
This looks like a circular reasoning, but actually it is a progress for a theory in terms of economy of though and potential unification of different areas of physics (for instance unification of thermodynamics and fluids mechanics).

However, for the purpose of this paper two points are worth emphasizing.
The first is that, in the spirit of the mechanical approach, the independence of probabilities for the velocities of pre-colliding particles results from the impossibility to reach a sufficient accuracy about the initial conditions, i.e. it results from an incomplete knowledge (in this mechanistic conception, real stochasticity does not exist). The underlying conception of probabilities is therefore much closer to that of subjectivists than to that of objectivists (depiste a frequentist ambition). 

The second point is that, whatever its origin, that is to say either incomplete knowledge (usual meaning of chaos) or a true stochastic process, the molecular chaos results over time in a loss of correlation between microstates along the trajectory in the phase space. 
The system is no longer deterministic in the sense that the chain of causality, representing the trajectory of the system in the phase space, is broken.
The volume-preserving dynamics (Eq.\ref{volpres}) and the ergodic hypothesis can still be postulated but they have lost their principal physical justification and the corresponding inductive reasoning has lost its strength and is much more hypothetical.

\subsubsection{Poincaré-Zermelo recurrence paradox}

Here, comes into play the Poincar\'e recurrence theorem. Consider a system, with a bounded phase space $\Gamma$, that continuously undergoes over time a transformation $F$ that preserves the volume of any subset of $\Gamma$.
Hamiltonian systems obey this condition according to the Liouville's theorem, but here the condition is more general and can apply not only to deterministic but also to stochastic (purely random) systems like the Ehrenfest urn model\,\cite{Karlin_1965}. 
Then, Poincar\'e shows that the system will recurrently pass to any point of $\Gamma$ already visited. 

Going back to the example above of the free expansion of a gas from a small box into a larger room, yes the gas expands, but it is also expected to return in the box on its own, without any demon. 
Although it would take a long time, it is not impossible, not just once but recurrently.

Hence the paradox stated by Zermelo (for an historical perspective see \cite{Brown_2009}): How can such a recurrence be consistent with a continuously increasing H-function? How, also, can it be consistent with a stable state of equilibrium?

Different arguments have been put forward to resolve the Poincaré-Zermelo paradox.
For instance, no system has a strictly bounded phase space, even the universe is expanding. 
Or, in the thermodynamical limit of an infinite number of particles, the time of recurrence is also infinite.
The argument initiated by Boltzmann himself is that for concrete thermodynamic systems with a very large number of particles, the calculated average time of recurrence is greater than the age of the universe. Practically, the gas never returns to the box on its own. So that everything is a question of time-window: \myquote{The range of validity of Boltzmann’s equation ... is limited in time by phenomena such as the Poincar\'e recurrence}(Villani\,\cite{Villani_2012}), but this limitation is never reached.

All these arguments are valid for resolving the paradox, they all amount to saying that in practice there is no recurrence.
But it remains a problem. If there is no recurrence, how to conceive probabilities as frequencies?
Or in the reverse manner, if there is recurrence, how to reconcile it with the 2nd law (H-theorem) and with a stable equilibrium?
The probabilist inference offered by information theory avoids this inconsistency.

%The answer is quite simple: abandon \textit{a posteriori} probabilities in favor of prior probabilities.

\subsection{Demons}

Demons observe thermodynamic systems, acquire information about them, and use it to act on them. In doing so they can possibly produce energy. Where does this energy come from?
In fact, energy is an abstraction only defined by a conservation principle\,\cite{Feynmann_Energy}. So that, if something is missing in an energy balance, it means that we have discovered a new form of energy. Demons, in their own, demonstrate the link between information and energy.
The same idea can be expressed in another manner: \myquote{In so far as the Demon is a thermodynamic system already governed by the Second Law, no further supposition about information and entropy is needed to save the Second Law.} Earman \& Norton\,\cite{Earman_1999}. In other words, the very definition of a principle is that everything conforms to it, by definition a principle is inviolable.

This is an application of pure logic with which I completely agree.
However, given the great expenditure of gray matter devoted to this question, such an answer cannot suffice. \myquote{How does it happen that there are people who do not understand mathematics? If the science invokes only the rules of logic, those accepted by all well-formed minds, if its evidence is founded on principles that are common to all men, and that none but a madman would attempt to deny, how does it happen that there are so many people who are entirely impervious to it?} (Poincar\'e\,\cite{Poincare_1914_en} p.46).
In fact, demons raise paradoxes that exist, and continue to exist even after they have been \textquote{resolved}, by the mere fact that they have been stated. So that we cannot shrug them off only by pure logic, we need more. 
Thus in what follows, we do not use the shortcut given in the preamble and rather examine whether what we know from the encoding problem is sufficient for understanding how demons can operate in accordance with the 2nd law.
No further supposition about information and entropy is needed to save the 2nd law\,\cite{Earman_1999}, but we suppose that we already have information theory at our disposal to solve the inconsistencies raised in the previous section. So that here, we just want to check its consistency with demons.

With the encoding problem, the quantity of information needed to represent a system, or equivalently the uncertainty about its state, is identified with its entropy, thus linked to energy.
In particular, acquiring information, i.e. reducing uncertainty, requires an energy expenditure (Eq.\ref{cineq2}-\ref{Wbit1}).
This acquired information is similar to potential energy, it is stored and could be used in return. Increasing the potential energy requires providing work, but to use it in return, something else is needed: know-how.
Otherwise the potential-information-energy is simply dissipated at the end of a cycle, that is when the information is outdated.
In short, it is not the acquisition of information which directly has an effect on the system, it is the acquisition plus the action which depends on it.
A misunderstanding of this point is at the origin of ill-founded criticisms of information theory (e.g.\,\cite{Callender_2004}).

Demons are supposed to know how, but the realization requires a physical implementation, not only of the action itself but also of all the information processing chain.
By physical implementation of information processing, I mean a black box including everything necessary for measurement, storage, transmission, eventual erasure, etc, that necessarily falls under the 2nd law of thermodynamics. It is this physical implementation that is responsible for the minimum energy expenditure for demons to operate.

From the 2nd law of thermodynamics expressed in terms of information (Eq.\ref{cineq2}-\ref{Wbit1}), the acquisition of 1 bit of information has a minimum energy cost equal to $T \ln 2$. 
Therefore, to check if demons work in accordance with the 2nd law, it is enough to check if the quantity of information necessary for their action is consistent with the energy that can be obtained in return.

\subsubsection{Maxwell's demon}

The family of thermodynamical demons\,\cite{Rex_2017} was born with the temperature-demon of Maxwell\,\cite{Maxwell_1872}.
Imagine a gas in an insulating container separated in two compartments $A$ and $B$ by a thermally insulating wall in which there is a small hole.
A demon 
\myquote{can see the individual particles, opens and c1oses this hole, so as to allow only the swifter particles to pass from A to B, and only the slower ones to pass from B to A. He will thus, without expenditure of work, raise the temperature of B and lower that of A in contradiction to the second law of thermodynamics.} (\,\cite{Maxwell_1872} p.308).
The temperature difference between the two compartments can eventually be used for running a thermodynamic cycle and producing work.

A simplified version is the pressure-demon: particles whatever their speed can only pass from $A$ to $B$. This results in a pressure difference between the two compartments, which can be used for producing mechanical work.
Alternatively, in this simplified version the demon can be replaced by a concrete device, either by a one-way valve as proposed by Smoluchowski\,\cite{Rex_2017}, or by a ratchet-pawl mechanism\,\cite{Feynmann_Ratchet}, or by an electric diode and the gas particles by electrons\,\cite{Brillouin1950, Brillouin_1956_book}, then if the two compartments communicate by an additional channel, the device is expected to rectify thermal fluctuations and produce a net current of particles, which can deliver useful energy.
For the last two concrete devices it has been experimentally shown that they can work, provided that the rectifier (ratchet-pawl or diode) is cooled at a lower temperature than the rest of the system\,\cite{Bang_2018, Gunn:1969wr} in return for the entropy decrease.
Demons work in the same way once physically implemented.

\begin{figure}[!htbp]
	\begin{center}
		\includegraphics[width=1\linewidth]{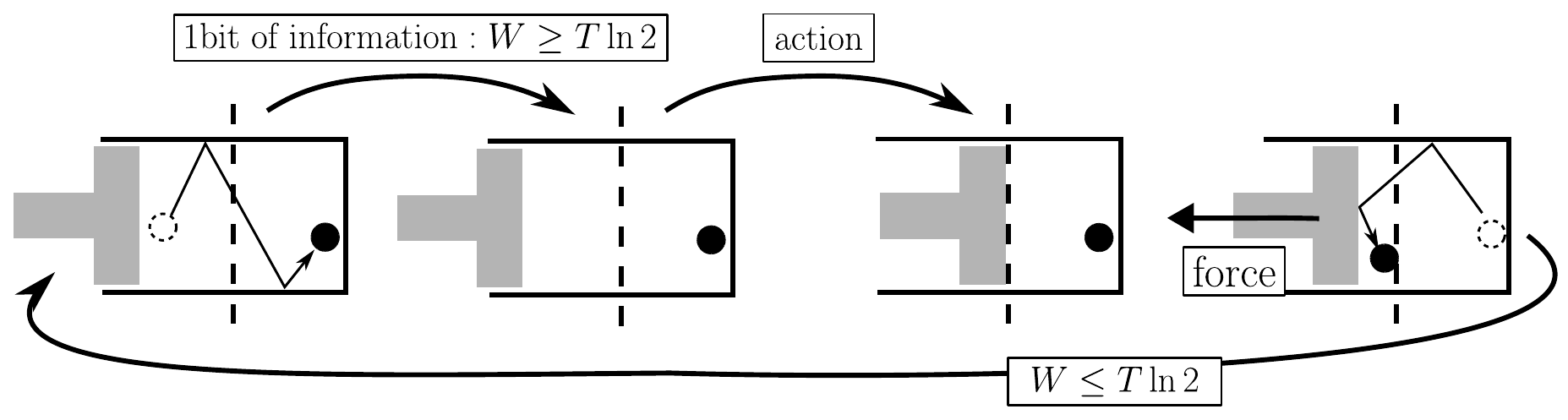}
		\caption{Szilard demon detects when the particle is in the suitable compartment and then installs a piston, allowing the device to subsequently produce work.}
		\label{szilard}
	\end{center}
\end{figure}

The quantitative verification of the correspondence between, on the one hand, the information necessary for the demon and, on the other hand, the energy that can be obtained in return is simplified with the device proposed by Szilard\,\cite{Szilard_1964}. It is composed of a single particle in a volume $2V$ (see Fig.\ref{szilard}). The demon does not care about the velocity of the particle but only the compartment it is in. This information is encoded with only 1 bit which costs at least $T\ln 2$ (Eq.\ref{Wbit1}).
In the opposite compartment, the demon installs a piston which encloses the particle in a volume $V$.
The system can return to the original state by an isothermal expansion that provides to the surroundings a work equal to at best $T\ln 2$. The overall cycle is consistent with thermodynamics.

\subsubsection{Loschmidt's demon}\label{Loschmidt_demon}

Let us return to the Loschmidt's paradox of reversibility (\S\ref{reverse}), this time with an operating demon actually able to reverse the velocities of particles. We wonder whether the quantity of information (in terms of the minimum number of bit to encode it) necessary for the demon to operate is in agreement with the mechanical work that the gas could produce with a subsequent isothermal expansion.

Let $V_1$ and $V_2$ denote respectively the volume of the initial box and that of the room in which the gas was expanded and $\lambda$ such as ${V_2}/{V_1}=2^\lambda$. So that the mechanical work per particle provided by an expansion is equal to at best $\lambda T \ln 2$, and for $N$ particles:
\begin{equation}
	W\le N \lambda T \ln 2
\end{equation}

For the demon to reverse the velocity of one particle in volume $V_2$, he must intercept the trajectory of the particle with an elastic wall (a mirror) having the correct direction (perpendicular to the trajectory), the correct orientation ($+$ or $-$), and the correct position (that of the particle)\,\cite{Binder_2023}. All the necessary information resides in the recording of the corresponding microstate of the particle.
From $V_1$ to $V_2$, the number of bit required for encoding the velocity of one particle remains unchanged, but encoding its position requires $\lambda$ extra bits (the cardinality of the phase space of one particle increases by a factor $V_2/V_1$). For $N$ particles, $N\lambda$ extra bits are needed.
From Eq.\ref{Wbit1} their acquisition costs at least $N\lambda T\ln 2$ in agreement with the work that can be obtained in return and with the 2nd law of thermodynamics.

\subsubsection{Landauer's \textquote{principle}}\label{shannonvslandauer}

Equation \ref{Wbit1}, obtained in the sole framework of Shannon's information theory (plus the 2nd law of thermodynamics), resembles strongly to another one known as Landauer's \textquote{principle}\,\cite{Landauer_1961, Landauer_1991, Bennett_1982, Bennett_2003} (that also uses the 2nd law but is free of Shannon's information theory and of the encoding problem).
Clarification is therefore necessary to avoid confusion.

Landauer considers the physical implementation of a logical bit under the form of a one-to-one mapping between the two logical states (0 and 1) and two thermodynamical states materialized for instance by a particle in a bistable potential. In this framework, it was shown that the irreversible logical operation ERASE (or RESET~TO~0) of the bit can be split into two steps:
\begin{enumerate}
	\item put the bit into an undetermined state by flattening the potential;
	\item set the bit to 0 by applying a bias and then raise back the energy barrier.
\end{enumerate}
The point is that during the first step the probability distribution of the particle undergoes a leakage comparable to the irreversible adiabatic free expansion of a gas (by a factor 2). So that neither heat nor work are exchanged with the surroundings. Whereas the second step can be quasistatic. Suppose that the initial state was 0, the second step closes a thermodynamic cycle. So that to be in agreement with the 2nd law, it must have an energy cost at least equal to $T\ln 2$. It follows that the total energy cost of the operation ERASE (of the cycle) is such as:
\begin{equation}\label{Landauer}
	W_\text{erase 1 bit} \ge T \ln 2
\end{equation}
This result is known as the Landauer's \textquote{principle} despite the fact that it cannot be a general principle (hence the quotes) but only applies to this particular physical implementation.
Actually, to avoid any leakage from one potential hole to its neighbor, it is enough to design a physical implementation based on a two-to-one mapping between logic and thermodynamic states\,\cite{Lairez_2023}. With only one potential hole there is no leak. 

Equation \ref{Wbit1} has general validity and concerns the acquisition of a data bit, whatever the way in which it was carried out and including all the steps necessary for this acquisition.
Equation \ref{Landauer} concerns the erasure of a data bit with a particular physical implementation consisting of a bistable potential and results from the thermodynamics of this particular case.
The difference between "acquisition" and "erasure" should be clear enough to avoid confusion. But, we can conceive certain particular data acquisition procedures (in particular that envisaged by Landauer and Bennet\,\cite{Landauer_1991, Bennett_2003} by which they propose to replace the thermostatistic demons) which require erasing the bit before writing a new value there. 
In this case equations \ref{Wbit1} and \ref{Landauer} lead to the same result. Hence a possible confusion.

The generic black box of the demons based on Eq.\ref{Wbit1}, that dissipates $T\ln 2$ per bit, includes everything necessary for measurement, storage, transmission, eventual erasure etc. Different physical implementations correspond to different places where dissipation could occur. There is absolutely no clue allowing to suggest that this place is universal.
Brillouin analyzed the physical limits for an observation through numerous examples of measurement procedure that could be implemented\,\cite{Brillouin_1956_book}. He showed that in all cases the energy expenditure to reach a given accuracy and the corresponding decrease of entropy that this information would allow are consistent with the 2nd law of thermodynamics.

\begin{figure}[!htbp]
	\begin{center}
		\includegraphics[width=1\linewidth]{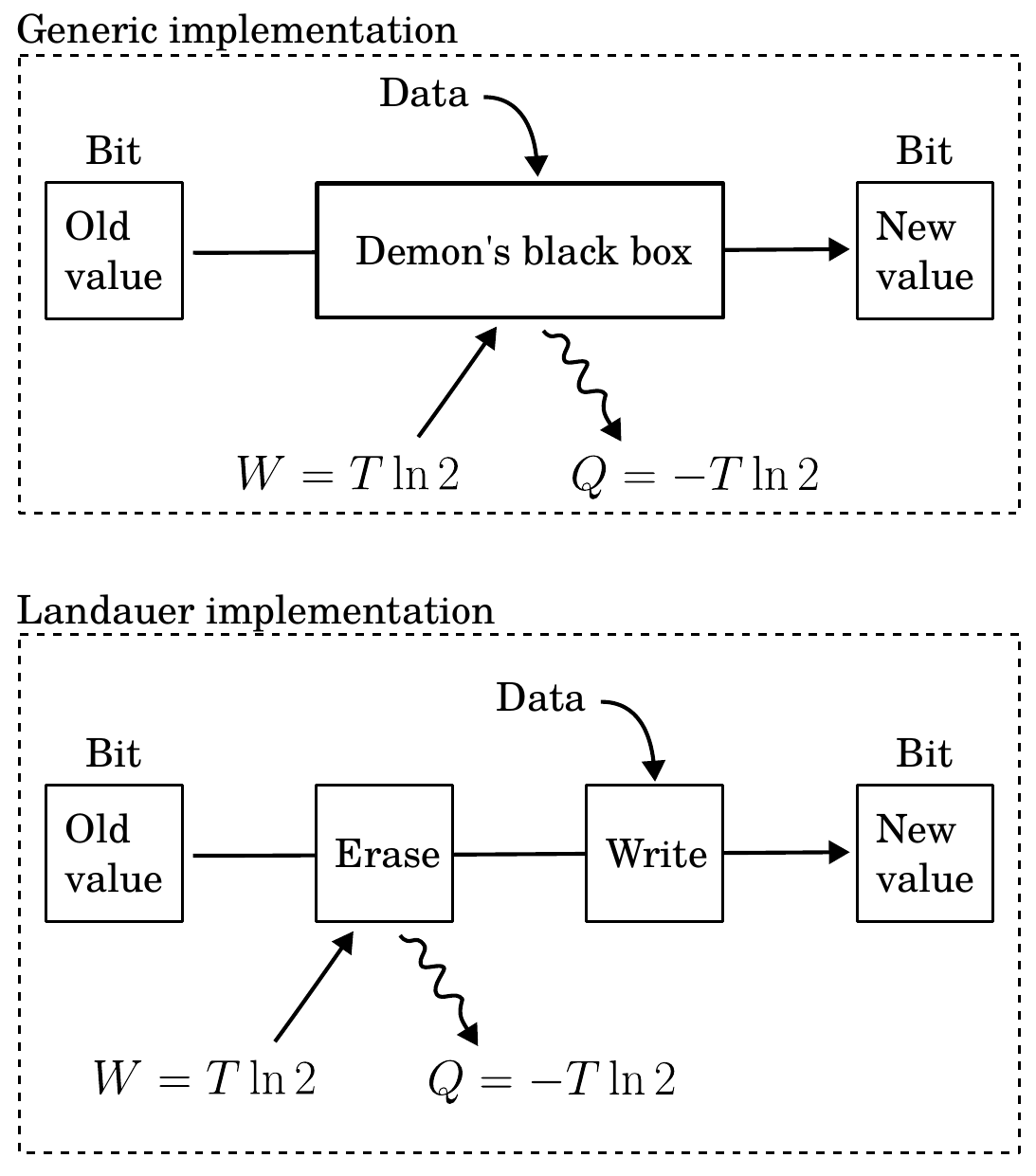}
		\caption{Generic implementation (Eq.\ref{Wbit1}) versus Landauer implementation (Eq.\ref{Landauer}) of a demon.}
		\label{Shavsland}
	\end{center}
\end{figure}

But currently, the most popular physical implementation is that of Landauer. 
The functioning of the Landauer's black box is such as the measurement is free and only its recording under the form of bit-values causes energy dissipation. For one bit of data, this functioning is as follows (see Fig.\ref{Shavsland}):
\begin{enumerate}
	\item The bit is materialized with a bistable potential. Erasing the bit dissipates at least $T\ln 2$ (Eq.\ref{Landauer}).
	\item The recording procedure requires the bit to be erased before new data is recorded.
\end{enumerate}
This functioning is a doubly special case: a special case of bit implementation and a special case of recording procedure.
Based on the 2nd law, it is obviously in full agreement with Eq.\ref{Wbit1}. So that, if it is claimed that a solution using Landauer's \textquote{principle} is found for the paradoxes introduced by demons; then, the same solution is valid using the sole framework of Shannon's information theory and Eq.\,\ref{Wbit1}. But this time in a more direct way with general validity.

Landauer's \textquote{principle} is presented as the key point for definitively resolving the paradoxes caused by demons, therefore to definitively link information to energy\,\cite{Berut_2012, Berut_2015, Ciliberto_2018, Yan_2018, Proesmans_2020, Giorgini_2023, Binder_2023, Oriols_2023}.
In addition to the previous objection that it is not a general principle, let me focus on the second point of the functioning of Landauer's black box for demons.

When I was a teenager I had a boombox to record my favorite music.
With this device it was possible to fully erase an already used cassette to start with an almost blank tape (a standard state). 
The idea behind this was that if you leave a blank between two pieces of music, you will not hear the old music when listening to the new.
But erasing the cassette was not mandatory. The cassette could be directly overwritten, for example if a long concert was recorded. In this case, the silence between two pieces (as silences within a given piece) is not a blank (an absence of message) but a message in itself. In other words, it is possible to record and process data without having to erase anything (the fact that the data are digitized or not, does not change anything). The injunction to avoid overwriting (and thus for the need of erasing), which we find in recent literature (e.g. \cite{Binder_2023}) is unfounded.

Actually, ERASE and OVERWRITE are irreversible logical operations. The logical irreversibility is defined by Landauer himself:
\myquote{We shall call a device [an operation] logically irreversible if the output of a device does not uniquely define the inputs.} \cite{Landauer_1961}.
It is a property of the initial and final states of the bit. For instance, a logical operation from $A$ to $B$
is logically irreversible if once in $B$ the information of where the system was initially has been lost, so that it is not possible to return to the starting point $A$.
On the contrary, a transformation is thermodynamically irreversible if it is not possible to return to $A$ by using the same path backward.
Thermodynamical irreversibility is a path property.
Therefore, it is not surprising that a logical irreversible operation can be achieved by using a reversible thermodynamic process.

\section{About subjectivity in physics}\label{subphy}

The approach of information theory to statistical mechanics avoids all inconsistencies of the alternative frequentist position.
But, we are forced to note an opposition to this idea.
This opposition goes beyond science and is epistemological.
The subjectivity introduced by information theory is the sticking point. 
Among many quotes, I note these which are particularly clear in this regard:

\myquote{The Jaynes approach [that of maximum entropy principle] is associated with a philosophical position in which statistical mechanics is regarded as a form of statistical inference rather than as a description of objective physical reality.} (Penrose\,\cite{Penrose_1979})

\myquote{A number of scientists and information theorists have maintained that entropy is a subjective concept and is a measure of human ignorance. Such a view, if it is valid, would create some profound philosophical problems and would tend to undermine the objectivity of the scientific enterprise.} (Denbigh \& Denbigh\,\cite{Denbigh_1985})

\myquote{This [the maximum entropy principle] is an approach which is mathematically faultless, however, you must be prepared to accept the anthropomorphic nature of entropy.} (Lavis\,\cite{Lavis_2015})

The aim of this section is to show that the type of subjectivity introduced into physics by information theory is in reality not new at all. 
It is in line with an ancient development of a general conception of what science is, which I propose to clarify.

\subsection{Representationalism}

The first subjective character introduced by information theory lies in the problem of coding a representation of the system which allows its behavior to be reproduced.

This representation depends on our knowledge, in the sense that it depends on the state of the art of the devices used to measure and probe the parameters needed for this representation. It also depends on the parameters that we consider relevant for this representation.
Consider for example the unavoidable impurities in any chemical substance. A correct representation of the system must take them into account. But below a given threshold which depends on our tools of measurement, impurities are no longer detectable and cannot be part of our representation of the system.
But we can suppose them to be still present in an objective being of the system.
Impurities are present in the real system, but not in our representation of it.
Additionally, impurities can be isotopes. What about the representation of the system before their discovery?\cite{Jaynes1992}. 
Also, in the thermodynamics of motors, pumps etc., most of the time we do not care about isotopes (actually we do not care about atoms either, before their discovery thermodynamic engines already worked very well), so that their presence or not is a useless information for the representation of the system (that does not need to be encoded).
In this sense, information is subjective. It depends on the state of knowledge of the observer or on which level of details we (collectively) consider as relevant to describe a system. But it does not depend on the personality of the observer, it is not a personal element\,\cite{Jaynes_1968}.

The subjectivity of entropy was already acknowledged by Maxwell (\myquote{The idea of dissipation of energy depends on the extent of our knowledge}\,\cite{Maxwell_1878})
and Gibbs (\myquote{It is to states of systems thus incompletely defined that the problems of thermodynamics relate.}\,\cite{Gibbs1874}).
This is believed to contrast with other physical quantities considered as objective properties. 
Actually, information theory does nothing other than an explicitly introduction of representationalism (also named indirect-realism) into these problems.

The basic idea of indirect-realism is that our only access to reality is that provided by our senses (in a broad sense that includes all laboratory instruments).
Following Einstein, \myquote{all knowledge about reality begins with experience and terminates in it}\,\cite{Einstein_1934}. If \textquote{experience} is understood as a conscious event that passes through our senses, it follows that the concern of science is not the reality but the representation our senses give us of it.
Fifty years before Shannon, Mach wrote:
\myquote{The law always contains less than the fact itself, because it does not reproduce the fact as a whole but only in that aspect of it which is important for us, the rest being either intentionally or from necessity omitted.}\,\cite{Mach_1898}. He was not talking about entropy or thermodynamics, he was talking about the laws of physics in general. 

Information theory formalizes this idea that would otherwise remain unclear and implicit.
Entropy itself is a very objective property well defined mathematically.
But it is an objective property of a subjective representation of the reality.
In this, according to indirect-realism entropy does not differ from other physical quantities.
The above argument should be able to answer the question asked by some:
\myquote{Thermodynamic entropy is not different, in regard to its status of objectivity, from physical properties in general. How came it then that so many scientists have held, and still hold, the opposite opinion?} (Denbigh \& Denbigh\,\cite{Denbigh_1985} p.18)

Regarding all physical quantities other than entropy, from a purely scientific point of view the difference between direct and indirect-realism is just a question of vocabulary. 
For direct realists, science is directly about the real world. 
For indirect realists, we do not have access to the real world but only to a representation of it, so that only this representation is the subject matter of science.
This difference can be ignored for a daily practice of science.
Rename \textquote{representation of reality} to \textquote{reality} and both realisms are talking about the same thing.
In both frameworks, theories concern the same object and their experimental examinations are equally achieved from our interactions with this object. 
The difference is only epistemological.
Poincar\'e wrote:
\myquote{Does the harmony [the laws of nature] the human intelligence thinks it discovers in nature exist outside of this intelligence? No, beyond doubt, a reality completely independent of the mind which conceives it, sees or feels it, is an impossibility. A world as exterior as that, even if it existed, would for us be forever inaccessible. But what we call objective reality is, in the last analysis, what is common to many thinking beings, and could be common to all; this common part  can only be the harmony expressed by mathematical laws. It is this harmony then which is the sole objective reality, the only truth we can attain.} (\cite{Poincare_1907} p.15).

When it comes to entropy as seen by information theory, what becomes troubling is that the difference between the direct and indirect-realism views can no longer be ignored, even in science.
 This is the only point on which entropy is so special compared to all other physical quantities.
% Because entropy makes the representation explicit.
Shannon entropy quantifies the complexity of the representation itself (in this case quantity of information is a particular case of complexity\,\cite{Grunwald_2004}). Doing so, it makes the notion of representation crucial and explicit.

\subsection{Induction}

Science is linked to knowledge, understood as a set of statements recognized as true.
According to  logical empiricism (or logical positivism), we have two possible sources of knowledge, each linked to a type of reasoning to assert that a statement is true: the first is purely logical (deductive reasoning) and the second empirical (inductive reasoning). 
Here, we will not enter into the debate on the justification of deduction, that is to say on the origin of the elementary rules of natural logic (which can possibly be empirical), we will focus only on induction, but we will need deduction for comparison.

Induction is unavoidable and omnipresent in natural sciences: generalization, interpolation, regression analysis, analogy etc., all are inductions based on known experimental facts. \myquote{Without generalisation, prediction is impossible} (Poincar\'e\,\cite{Poincare_1905_Sci_Hypo}). 
In fact, induction is the reasoning that allows us from our current knowledge to predict new observations or answer new questions.
At the basis of phenomenological laws, but also theoretical hypothesis, postulates or principles, there is always inductive reasoning, at least implicit.

However, if the truth of a deductive statement can be proved and verified (provided the premises and the logical rules are right), the verification of an inductive statement can never be definitively achieved because this would imply a infinite non-countable set of experimental facts. The truth of a deduction is certain, that of an induction is at best probable.
\myquote{By generalization, every fact observed enables us to predict a large number of others; only, we ought not to forget that the first alone is certain, and that all the others are merely probable} (Poincar\'e\,\cite{Poincare_1905_Sci_Hypo}).
If known experimental facts make it possible to base an inductive reasoning, new or upcoming ones can only either confirm or refute it, but never definitely prove it.
Inductions are by essence provisional and likely to be updated or replaced by better ones as progress is made.

If an inductive statement can never be verified (proven to be true), how can we make the difference between a well-founded scientific claim and another ill-founded and irrational? How can we make a hierarchy between different reasonings? What is the best?
This is known as the problem of induction (for a recent book on this topic see \cite{Huber_2019}).

A first piece of answer was provided by the falsificationism of Popper\,\cite{Popper_2005}. Since verifiability cannot be required for induction, Popper instead suggests replacing it with falsifiability.
A valid inductive reasoning must be falsifiable (or refutable): it must be able to be confronted with the experiments. This is the first condition, if it is met, an induction remains \textquote{true} until proven otherwise.
The requirement of falsifiability of an induction entails another, that of not being tacit or hidden, but explicit. Otherwise, we make them without any chance of attempting to refute them\,\cite{Poincare_1905_Sci_Hypo}.
But this is still not enough to establish a hierarchy of inductive reasoning.

Confirmation or refutation of inductive reasoning passes though experiments. At first glance, the refutation seems clear-cut, while the confirmation seems gradual (incremental) as more and more facts consistent with an induction reinforce it.
However, both are conditioned on the validity of experimental results, themselves conditioned on confidence intervals (errors bars). 
This automatically introduces a link between the notion of \textquote{degree of confirmation} or \textquote{degree of belief} and that of \textquote{probability of truth} of an induction\,\cite{Sprenger_2016}.
Hence the claim:
all inductive reasoning in science falls under the same universal pattern as that of probabilistic inferences.
The best is the most probable according to our present knowledge, that is to say the one having the highest prior probability of be true.
This is the essence of Bayesianism\,\cite{Sprenger_2019} (named after Bayes and his theorem about the probability of an event conditioned on prior knowledge) and its derivatives in spirit, among which the maximum entropy inference can be classified.

Not everyone agrees with the existence of such a universal pattern for induction. 
For example, Norton\,\cite{Norton_2021} introduces a material theory of induction which professes that the logic of induction is determined by facts specific to each case and which cannot always be expressed in terms of probability. To which it has been opposed\,\cite{Dawid_2015} that as soon as the confirmation procedure (and then the updating of the induction) involves data and measurements, probabilities come into play.

Whether maximum entropy inference is a starting point to produce a first prior probability distribution necessary for Bayesian updating, or whether it is a generalization of Bayesian inference, unless it is the other way around\,\cite{Jaynes_1988a}, is beyond the scope of this paper.
The main point is the universal aspect of all inductive reasoning, that of being ultimately probabilistic, that of involving prior knowledge and prior probabilities, that of being  subjective (in the sense given to this word in this paper).

Although scientists are aware of the problem of induction and adopt probabilistic inductive reasoning for their daily practice (personally, I don't know any scientist who would prefer to work on the option she believes has the lowest probability of success), this practice is not necessarily conscious and the problem of induction is often (temporarily) forgotten or denied. Below are some quotes from the recent literature of interest here:
\myquote{Experimental verification of Landauer’s principle linking information and thermodynamics.} (B\'erut et al.\,\cite{Berut_2012}).
\myquote{Information and thermodynamics: experimental verification of Landauer's Erasure principle.} (B\'erut et al.\,\cite{Berut_2015}).
\myquote{We	experimentally demonstrate a quantum version of the Landauer principle.} (Yan et al.\,\cite{Yan_2018}).
\myquote{Landauer’s principle has been recently verified experimentally} (Binder\,\cite{Binder_2023}), etc.
I cannot imagine that these authors ignore or disagree with the impossibility to verify an induction. Instead, I interpret these quotes as language facilities that are not innocent but reveal a reluctance to inductions.
Physicists prefer deductions, proofs and definitive verifications, all things expected of hard-sciences.

The problem with information theory is that here again, as for representationalism, everything is explicit: we cannot feign ignorance of our complete dependence on subjective probabilities in natural sciences.

\section*{Conclusion}

The subjectivity of information theory, as it has been defined in this paper, that is to say something which is not personal but simply refers to the role played by our knowledge, allows us to resolve the inconsistencies present in thermostatistics from the start. At the same time, it is this subjectivity that worries some for epistemological rather than scientific reasons.

The role played by subjectivity should not be so surprising, at least in this area.
Thermodynamics from the beginning refers to anthropocentric concepts and vocabulary such as energy grades, useful energy, energy cost, work, dissipation...
In addition, thermodynamics is a science of the macroscopic scale. This term itself is anthropocentric, since macroscopic only designates our human scale.
Indeed, in practice, the role that a certain subjectivity can play is admitted in science. But we are so steeped in positivism and with the ambition to be objective, that when subjectivity becomes too explicit it becomes annoying.
In fact, science is a human construct. The \textquote{Laws of Nature} do not come from Nature, they come from us.
The mere fact that these laws are provisional and subject to being continually replaced (updated) by better ones as science and our knowledge progress proves this.

Finally, there is another source of reluctance towards information theory that can be perceived in light of certain recent publications. It was not mentioned in this article but probably deserves special attention. It is also linked to the ambition of objectivity, but not in the same way as the refusal of representationalism and Bayesianism was. It is due to a particular meaning given to the word \textquote{physical}, understood as \textquote{materialized}, as opposed to virtual or non-tangible.
There is nothing more \textquote{objective} than matter.
After Landauer, (\textquote{Information is physical}\,\cite{Landauer_1991}), many authors interpreted his principle as the missing element they were waiting for to materialize information. 
Probably the most recent development of this idea can be found in the \textquote{mass-energy-information equivalence principle}\,\cite{Vopson_2019, Vopson_2022}. In short: information is energy, energy has a mass equivalent in special relativity ($E=mc^2$), therefore information would have a mass. For example, the author proposes measuring the mass of a hard drive before and after erasing 1\,TB of data.
The idea behind it is that information is a kind of potential energy, which can be used for instance by a demon to act on a system and produce work.
What is the status of potential energy in special relativity?\,\cite{Brillouin_1965a,Brillouin_1965b}. 
Actually, potential energy is not energy, it is something that is potentially energy. 
It is actually stored under the form of rest-mass\,\cite{Hecht_2016}, just like the mass defect in nuclear physics. Hence the idea behing the \textquote{mass-information equivalence principle}.
Potential energy is a concept, just like entropy and information are. But there is a major difference: if we wonder about the mass equivalence of $T\Delta S$ when a body undergoes a process, we implicitly consider a constant temperature and therefore a constant internal energy to which no variation in rest mass can be associated as being localized in the body considered, but more likely in its surroundings. In this sense, entropy and information remain even more elusive concepts than that of potential energy. But developing this point deserves another article\,\cite{Lairez_2024a}.

\bibliography{\string~/Documents/Articles/weri_biblio.bib}

\end{document}